\documentclass{apex}
\usepackage{graphicx}
\usepackage{dcolumn}
\usepackage{bm}

\title
{
Epitaxial Growth of FeSe$_{0.5}$Te$_{0.5}$ Thin Films on CaF$_2$ Substrates 
with High Critical Current Density
}

\author{
Ichiro TSUKADA$^{1,5}$\thanks{E-mail address: ichiro@criepi.denken.or.jp}, 
Masafumi HANAWA$^{1,5}$, 
Takanori AKIIKE$^{2,5}$, 
Fuyuki NABESHIMA$^{2,5}$, 
Yoshinori IMAI$^{2,5}$, 
Ataru ICHINOSE$^{1,5}$, 
Seiki KOMIYA$^{1,5}$, 
Tatsuo HIKAGE$^{3,5}$, 
Takahiko KAWAGUCHI$^{4,5}$, 
Hiroshi IKUTA$^{4,5}$, 
and Atsutaka MAEDA$^{2,5}$
}

\inst{
$^{1}$Central Research Institute of Electric Power Industry, 2-6-1 Nagasaka, Yokosuka, Kanagawa 240-0196, Japan \\
$^{2}$Department of Basic Science, The University of Tokyo, 3-8-1 Komaba, Meguro, Tokyo 153-8902, Japan \\
$^{3}$High Intensity X-ray Diffraction Laboratory, Nagoya University, Chikusa, Nagoya 464-8603, Japan \\
$^{4}$Department of Crystalline Materials Science, Nagoya University, Chikusa, Nagoya 464-8603, Japan \\
$^{5}$JST, TRIP, Sanbancho, Chiyoda, Tokyo 102-0075, Japan
}

\abst{
{\it In-situ} epitaxial growth of FeSe$_{0.5}$Te$_{0.5}$ thin films is demonstrated 
on a nonoxide substrate CaF$_2$. 
Structural analysis reveals that compressive stress is moderately 
added to 36-nm-thick FeSe$_{0.5}$Te$_{0.5}$, 
which pushes up the critical temperature to above 15~K, 
showing higher values than that of bulk crystals. 
The critical current density at $T$ = 4.5~K reaches 
5.9 $\times$ 10$^4$~Acm$^{-2}$ at $\mu_0H$ = 10~T, 
and 4.2 $\times$ 10$^4$~Acm$^{-2}$ at $\mu_0H$ = 14~T. 
These results indicate that fluoride substrates have high potential for the growth 
of iron-based superconductors in comparison with popular oxide substrates.
}

\begin{document}
\maketitle

%(Introduction1)

Since the discovery of iron-based superconductors,
\cite{Kamihara1} 
much effort has been devoted to establish a thin-film growth technique 
of these compounds.
\cite{Hiramatsu1,Hiramatsu2,Backen1,Iida1,Kawaguchi1,Kawaguchi2,Takeda1} 
At an early stage, iron-chalcogenide superconductors
\cite{Hsu1} 
had been considered rather inappropriate for practical applications 
simply because of their low critical temperature ($T_c$). 
However, pressure-effect studies demonstrate 
that the potential $T_c$ of iron-chalcogenide superconductors is as high as 37~K,
\cite{Margadonna1,Medvedev1} 
which motivated us to begin the study of thin-film growth of 
FeSe$_{1-x}$Te$_x$. 
Through many reports, 
\cite{Wu1,Han1,Bellingeri1,Si1,Imai1}
one problem has gradually emerged; 
There is a close correlation between $T_c$ and the structure of the films 
grown epitaxially, 
but their lattice parameters are influenced by too many growth parameters, 
and the lattice parameters of the substrate material are not a dominant factor.
\cite{Bellingeri2,Imai2,Hanawa1} 
In other words, FeSe$_{1-x}$Te$_x$ can be grown epitaxially on 
a single-crystalline substrate, but the lattice parameters cannot be 
designed in accordance with those of the substrates. 
One of the possible reasons that we have previously proposed for this 
is oxygen contamination from oxide substrates,
\cite{Imai2,Hanawa1} 
and we actually confirmed the presence of oxygen at the interface 
of FeSe$_{1-x}$Te$_x$ grown on YSZ and LaSrAlO$_4$. 
In order to avoid this problem and further improve superconducting properties, 
it is important to investigate the film growth on nonoxide substrates. 
In this Letter, we demonstrate an epitaxial growth on nonoxide 
single-crystalline substrates, CaF$_2$ ($a_0$ = 5.463~{\AA}). 
Our results indicate that films with high-$T_c$ and high-$J_c$ can be 
obtained reproducibly on CaF$_2$ (100) substrates 
even with a thickness as small as approximately 40~nm. 
Several films showed a $T_c$ higher than that of bulk single crystals, 
strongly suggesting that the epitaxial strain effect moderately works 
on CaF$_2$ (100).

%(Film Growth)

All the films were grown by pulsed laser deposition from an FeSe$_{0.5}$Te$_{0.5}$ 
target, as was described elsewhere.
\cite{Imai1,Imai2,Hanawa1} 
Substrate temperature, laser repetition rate, and back pressure are 280$^{\circ}$C, 
10~Hz, and $\sim$ 10$^{-5}$~Torr, respectively. 
Commercially available CaF$_2$ (100) substrates were used for the present experiments. 
A specially designed metal mask was put directly on the substrate to make the films 
into a six-terminal shape, in which the dimensions of the measured area are 
0.95~mm length and 0.2~mm width. 
We prepared four very thin films with thicknesses of approximately 40~nm 
named C1 (36~nm), C2 (38~nm), C3 (40~nm), and C4 (42~nm), 
and two relatively thick films with thicknesses of approximately 150~nm 
named C5 (150~nm) and C6 (175~nm). 
The resistivity and critical current were measured using a physical property 
measurement system (PPMS) under magnetic fields of up to 14~T.

%(XRD: c-axis and in-plane) 

\begin{figure}
\begin{center}
\includegraphics*[width=85mm]{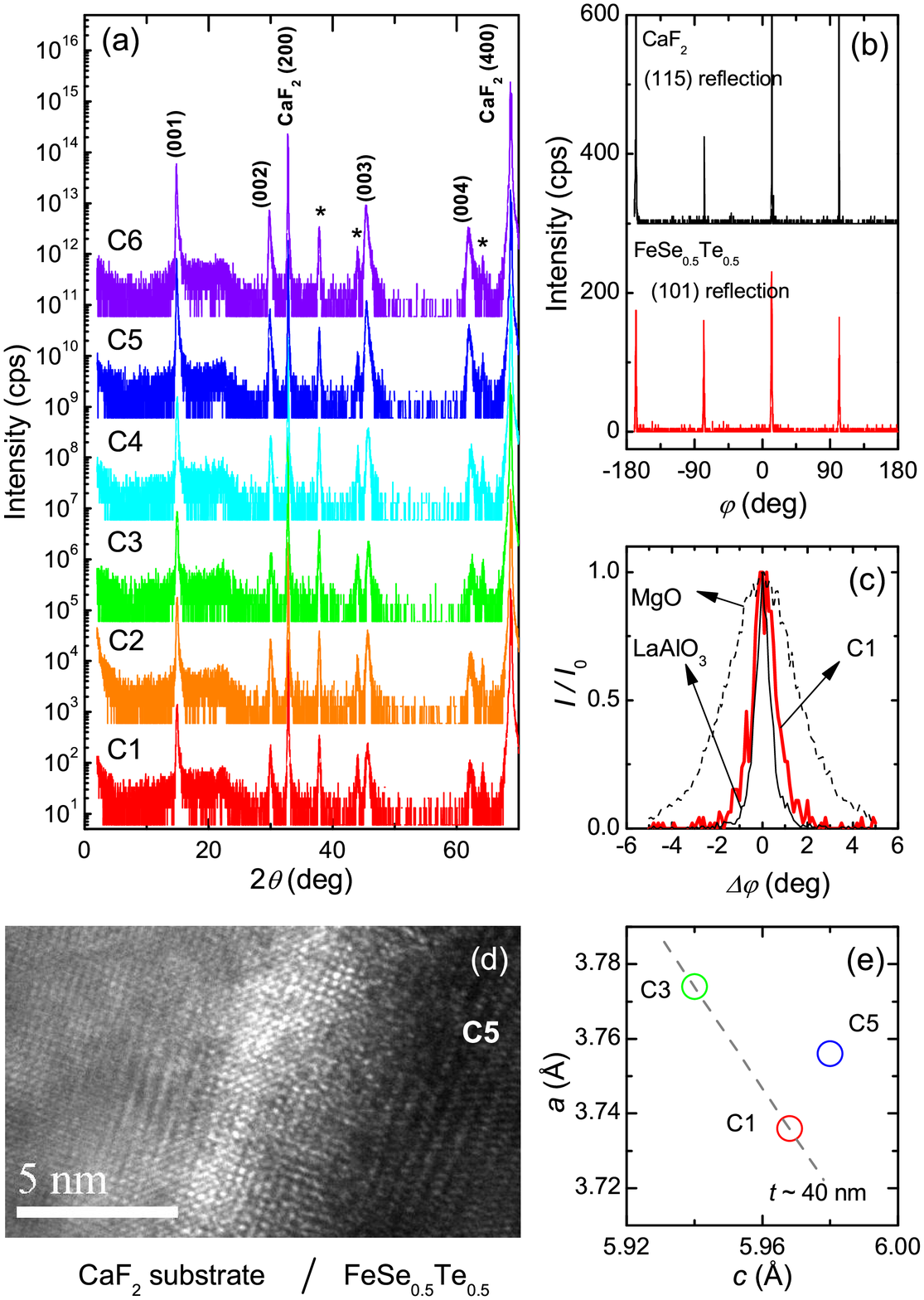}
\end{center}
\caption{(color online) 
(a) X-ray diffractions of six FeSe$_{0.5}$Te$_{0.5}$ thin films grown on 
CaF$_2$ (100). The origin of the vertical axis is shifted for each film. 
$\ast$ indicates peaks from an aluminum holder. 
(b) $\phi$-scans of the 101 reflection of film C1 
and the 115 reflection of CaF$_2$ substrate. 
(c) Comparison of the 101 reflections of film C1 
with the films on LaAlO$_3$ and MgO in Ref. 18).
(d) Cross-sectional TEM image at the interface of CaF$_2$ and film C5. 
(e) $c$-axis length vs $a$-axis length of films C1, C3, and C5. 
The dashed line is a guide to the eye.
}
\label{Fig.1}
\end{figure}

All the films are $c$-axis oriented as shown in Fig.~\ref{Fig.1}(a), 
and no reflections originating from an impurity phase are detected. 
In-plane orientation is confirmed for C1, C3, and C5 by a 4-circle diffractometer. 
Figure~\ref{Fig.1}(b) shows the $\phi$ scans of the 101 reflection 
of FeSe$_{0.5}$Te$_{0.5}$ and the 115 reflection of CaF$_2$ in film C1. 
A 4-fold symmetry reflection is obtained as observed in the films 
on several oxide substrates,
\cite{Imai2} 
and we can identify the in-plane orientation as FeSe$_{0.5}$Te$_{0.5}$ [100] 
$\parallel$ CaF$_2$ [110]. 
It should be noted that the peak width of the 101 reflection is comparable to 
but not better than that reported in the film prepared on LaAlO$_3$ 
as shown in Fig.~\ref{Fig.1}(c); 
FWHM of the peak is $\approx$ 1.0$^{\circ}$ for C1 
showing a larger value than reported for the film on LaAlO$_3$.
\cite{Imai2} 
Figure~\ref{Fig.1}(d) shows the cross-sectional TEM image of film C5 
at the interface. 
Roughness of the boundary looks to be more emphasized when compared with 
the case of LaAlO$_3$ substrate,
\cite{Imai2,Hanawa1} 
but no amorphous layer is observed between the film and the substrate.

The calculated $c$- and $a$-axis lengths are summarized in Fig.~\ref{Fig.1}(e); 
the former is calculated from position of the 001 - 004 reflections, and the 
latter is determined from the position of 101 and the calculated $c$-axis length. 
The $c$-axis lengths of all the films exceed 5.94~{\AA}, 
and are longer than that reported for the films grown on oxide substrates 
in our previous paper.
\cite{Imai2,Hanawa1} 
The $a$-axis lengths of three films show correspondingly short values, 
less than 3.78~{\AA}, 
which are comparable with those reported by Bellingeri {\it et al.}
\cite{Bellingeri2} 
Thus, it is concluded that these films are compressed along 
the $a$-axis and are elongated simultaneously along the $c$-axis. 
Note that such a short $a$-axis length of the film is not due to 
a simple coincidence to the lattice parameters of CaF$_2$ 
($a_0 / \sqrt{2}$ = 3.863~{\AA}). 
The reason why the $a$-axis is so strongly compressed has not yet been 
clarified, and we can only speculate on a possible chemical reaction 
between FeSe$_{1-x}$Te$_x$ and F and/or some unknown mechanisms 
in FeSe$_{1-x}$Te$_x$ to shrink the $a$ axis in oxygen-free environment.

%(RT1)

\begin{figure}
\begin{center}
\includegraphics*[width=80mm]{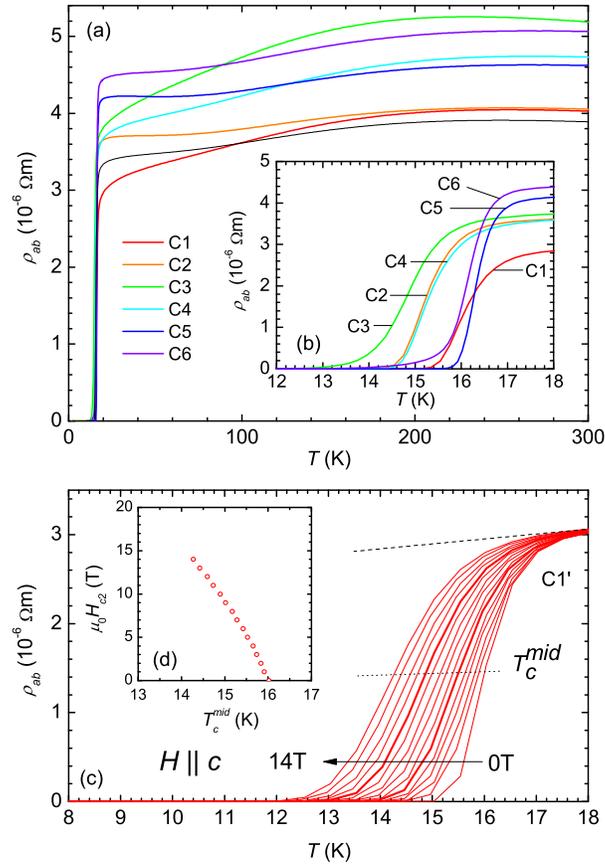}
\end{center}
\caption{(color online) 
(a) Temperature dependence of resistivity of six thin films grown on CaF$_2$, 
and (b) its closeup around $T_c$. 
(c) $\rho$-vs-$T$ measured for film C1' under magnetic fields up to 
$\mu_0H$ = 14~T applied along the $c$-axis.
\cite{C1'} 
The dashed line indicates normal-state resistivity, 
and the dotted line indicates the position of half of the normal resistivity. 
(d) $T_c^{mid}$-vs-$H_{c2}$ plot of film C1' . 
}
\label{Fig.2}
\end{figure}

The temperature dependences of resistivity of the six films are 
summarized in Fig.~\ref{Fig.2}(a). 
All the films exhibit a higher $T_c$ than those in our previous works.
\cite{Imai2,Hanawa1} 
$T_c$'s are concentrated within a narrow temperature region around 15~K 
with high reproducibility, even though the normal state resistivities 
are rather scattered. 
The observation of the less-scattered $T_c$'s shows a clear contrast 
to the results of the films on various oxide substrates.
\cite{Imai2,Hanawa1} 
Among the six films, C1 and C5 are remarkable for their $T_c$'s exceeding 14~K, 
which is the highest $T_c$ value reported for FeSe$_{1-x}$Te$_x$ single crystals 
in ambient pressure; 
the midpoint $T_c$ ($T_c^{mid}$) and zero-resistivity $T_c$ ($T_c^{zero}$) 
are 16.1 and 15.2~K for C1, and 16.3 and 15.3~K for C5, respectively. 
This is not the first example, and a higher $T_c$ (onset) of 21~K 
has been already reported by Bellingeri {\it et al.}
\cite{Bellingeri2} 
However, they have obtained such a high $T_c$ only in relatively thick films 
of 200~nm. 
We emphasize in the present study that even a far thinner film can 
exhibit a higher $T_c$ than bulk crystals, and we are sure that this is 
one of the significant benefits of using a CaF$_2$ oxygen-free substrate. 
In particular, when fabricating SNS Josephson junctions and SIS tunnel junctions, 
it becomes very important to prepare a sharp interface between superconductors 
and normal metals and/or barrier insulators. 
For that purpose, thinner film have an advantage of smoother surfaces than thicker films, 
and we believe that CaF$_2$ will work as one of the promising substrates.

These films show remarkable properties also in a magnetic field. 
Figure~\ref{Fig.2}(c) shows a suppression of $T_c$ by applying magnetic 
fields along the $c$-axis direction ($H \parallel$ $c$).
\cite{C1'} 
Even under 14~T, the decrease of $T_c^{mid}$ is less than 2~K, 
and thus the upper critical field ($H_{\rm c2}$) is suggested to be high. 
However, $H_{\rm c2}$ at the region close to $T_c$ is not linear to $T$ 
suggesting that an application of the WHH relation to the present case is invalid. 
With a simple application of the WHH relation to estimate $H_{\rm c2}$ 
using $H_{\rm c2}$(0) = - 0.69$T_c(dH_{\rm c2}/dT)\vert_{T = T_c}$, 
we obtained $H_{\rm c2}$ = 120~T. 
This is obviously overestimated, and the actual $H_{\rm c2}$ 
must be less than this value. 
However, even if we use a linear extrapolation of the tangential line 
at $\mu_0H$ = 14~T instead, and multiply 0.69 to the value at $T$ =0~K, 
we obtain $H_{\rm c2}$ = 74.5~T, which is still larger than those 
reported for FeSe$_{0.5}$Te$_{0.5}$ using $T_c^{mid}$,
\cite{Klein1} 
and even using $T_c^{onset}$.
\cite{Tsurkan1} 
This result demonstrates the robustness of superconductivity in our FeSe$_{0.5}$Te$_{0.5}$ 
thin films under magnetic fields.

%(Jc)

\begin{figure}
\begin{center}
\includegraphics*[width=85mm]{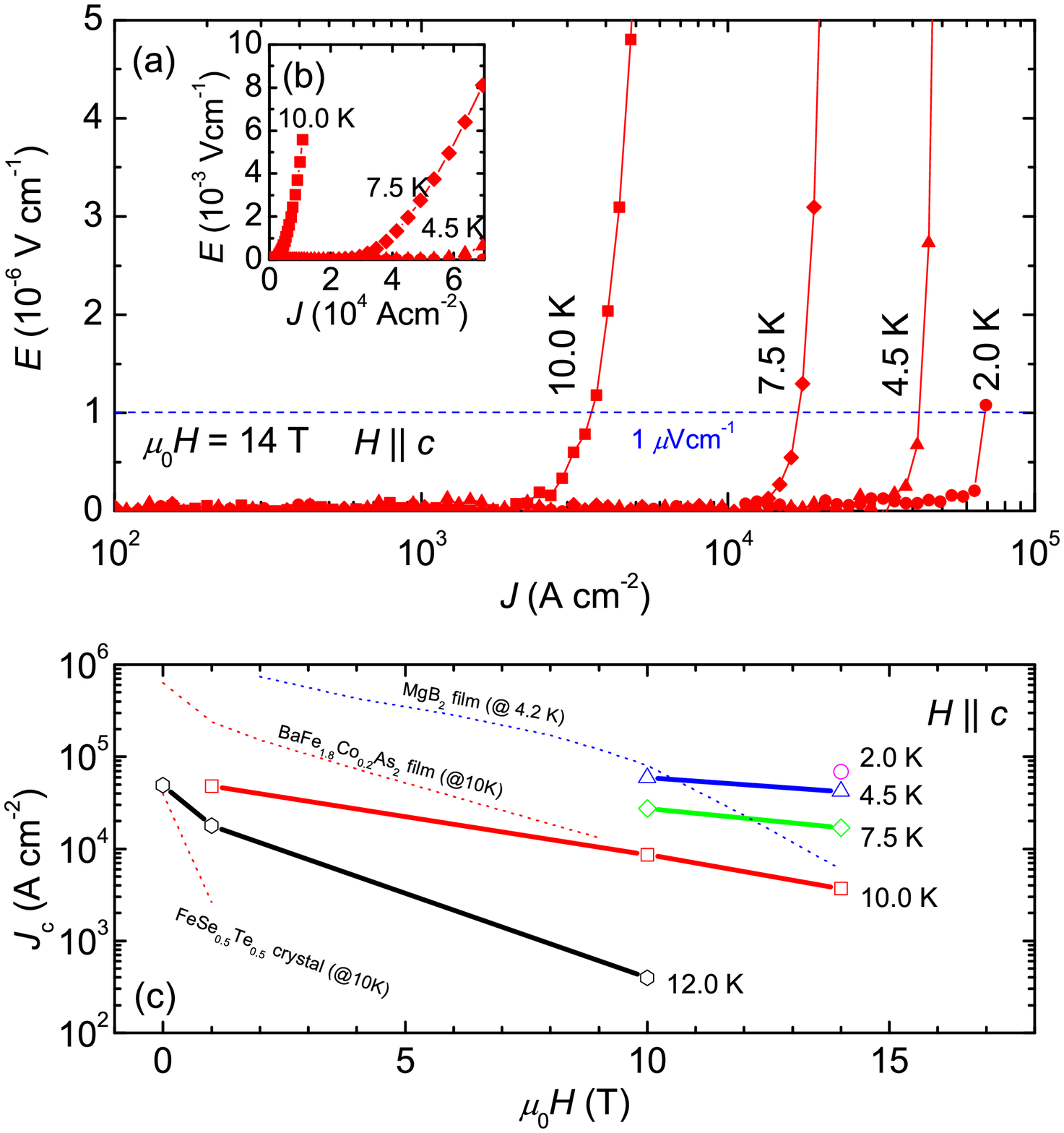}
\end{center}
\caption{(color online) 
(a) Current-density dependence of electric field in film C1. 
The maximum current available in PPMS is 5~mA corresponding to $J$ $\approx$ 7 $\times$ 10$^4$~Acm$^{-2}$. 
(b) Large-scale plot of the same data. 
(c) Applied-field dependence of $J_c$ at $T$ = 2.0, 4.5, 7.5, 10.0, and 12.0~K. 
The data for FeSe$_{0.5}$Te$_{0.5}$ crystal,\cite{Tsurkan1} 
BaFe$_{1.8}$Co$_{0.2}$As$_2$ thin film,\cite{Iida2} 
and MgB$_2$ thin film\cite{Eom1} 
are also plotted. 
}
\label{Fig.3}
\end{figure}

Finally, we evaluate the potential critical-current density of FeSe$_{0.5}$Te$_{0.5}$ 
for the application to thin-film conductors. 
The excellent properties observed in film C1 suggests that this film can show 
a high critical current density ($J_c$) in a superconducting state. 
We performed resistivity measurements with different excitation currents 
in the region below $T$ = 15~K at $\mu_0H$ = 0, 1, 10, and 14~T. 
Part of the data at $\mu_0H$ = 14~T in $H$ $\parallel$ $c$ configuration 
is shown in Figs.~\ref{Fig.3}(a) and \ref{Fig.3}(b). 
$J_c$ is defined as the current density at which a finite electric field of 
1~$\mu$Vcm$^{-1}$ emerges. 
The field-dependence of $J_c$ is summarized in Fig.~\ref{Fig.3}(c) 
at selected temperatures with the reported $J_c$ values on several materials. 
It is clearly seen that at $T$ = 10~K, $J_c$ of film C1 is one order 
of magnitude higher than that of FeSe$_{0.5}$Te$_{0.5}$ crystal,
\cite{Tsurkan1} 
and approaches the value reported by Iida {\it et al.} for 
BaFe$_{1.8}$Co$_{0.2}$As$_2$ thin films,
\cite{Iida2} 
even though the value is still lower than that reported by Lee {\it et al.}
\cite{Lee1} 
The most remarkable feature is that $J_c$ is suppressed very slowly above 10~T. 
At $T$ = 4.5~K, $J_c$ roughly equals 5.9 $\times$ 10$^4$~Acm$^{-2}$ at 
$\mu_0H$ = 10~T, and 4.2 $\times$ 10$^4$~Acm$^{-2}$ at $\mu_0H$ = 14~T. 
The latter value is larger than that of MgB$_2$ thin films 
by almost one order of magnitude,
\cite{Eom1} 
which suggests that iron-chalcogenide superconductor thin films have 
high potential under high-magnetic fields. 
It should also be noted that the present film is a pristine film 
simply prepared on CaF$_2$ without buffer layers. 
There is room for further improvement of $J_c$ by introducing an external 
pinning center to the film, which may push up the critical current density 
to above 10$^5$~Acm$^{-2}$ in high field regions.

%(summary)

To summarize, we have grown FeSe$_{0.5}$Te$_{0.5}$ thin films on a CaF$_2$ (100) 
substrate, and succeeded in obtaining films showing $T_c$ as high as 15~K 
with sufficiently high reproducibility. 
The structural analysis indicates that the films are compressed along 
the $a$-axis. 
The high critical current density is also demonstrated. 
The value of $J_c$ reaches 5.9 $\times$ 10$^4$~Acm$^{-2}$ at $T$ = 4.5~K 
and $\mu_0H$ = 10~T, which is comparable to that of Co-doped BaFe$_2$As$_2$ 
thin films. 
Better superconductivity has been recently confirmed in NdFeAs(O,F) thin films 
prepared on the same substrate material,
\cite{Ikuta1} 
which indicates the common advantage of using CaF$_2$ to grow iron-based superconductor 
thin films.

%(Acknowledgment)

We thank D. Nakamura for technical assistance.

\newpage

\begin{thebibliography}{9}
\bibitem{Kamihara1} Y. Kamihara, T. Watanabe, M. Hirano, and H. Hosono: 
J. Am. Chem. Soc. {\bf 130}, 3296 (2008). 
\bibitem{Hiramatsu1} H. Hiramatsu, T. Katase, T. Kamiya, M. Hirano, and H. Hosono: 
Appl. Phys. Lett. {\bf 93}, 162504 (2008). 
\bibitem{Hiramatsu2} H. Hiramatsu, T. Katase, T. Kamoya, M. Hirano, and H. Hosono: 
Appl. Phys. Express {\bf 1}, 101702 (2008). 
\bibitem{Backen1} E. Backen, S. Haindl, T. Niemeier, R. H{\"u}hne, T. Freudenberg, J. Werner, G. Behr, L. Schultz, and B. Holzapgel: 
Superconduct. Sci. Technol. {\bf 21}, 122001 (2008). 
\bibitem{Iida1} K. Iida, J. H{\"a}nisch, R. H{\"u}hne, F. Kurth, M. Kidszun, S. Haindl, J. Werner, L. Schultz, and B. Holzapfel: 
Appl. Phys. Lett. {\bf 95}, 192501 (2009). 
\bibitem{Kawaguchi1} T. Kawaguchi, H. Uemura, T. Ohno, R. Watanabe, M. Tabuchi, T. Ujihara, K. Takenaka, Y. Takeda, and H. Ikuta: 
Appl. Phys. Express {\bf 2}, 093002 (2009). 
\bibitem{Kawaguchi2} T. Kawaguchi, H. Uemura, T. Ohno, M. Tabuchi, T. Ujihara, K. Takenaka, Y. Takeda, and H. Ikuta: 
Appl. Phys. Lett. {\bf 97}, 042509 (2010). 
\bibitem{Takeda1} S. Takeda, S. Ueda, T. Yamagishi, S. Agatsuma, S. Takano, A. Mitsuda, and M. Naito: 
Appl. Phys. Express {\bf 3}, 093101 (2010). 
\bibitem{Hsu1} F. C. Hsu, J. Y. Luo, K. W. Yeh, T. K. Chen, T. W. Huang, P. M. Wu, Y. C. Lee, Y. L. Huang, Y. Y. Chu, D. C. Yan, and M. K. Wu: 
Proc. Natl. Acad. Sci. USA {\bf 105} (2008) 14262. 
\bibitem{Margadonna1} S. Margadonna, Y. Takabayashi, Y. Ohishi, Y. Mizuguchi, Y. Takano, T. Kageyama, T. Nakagawa, M. Takata, and K. Prassides: 
Phys. Rev. B {\bf 80}, 064506 (2009). 
\bibitem{Medvedev1} S. Medvedev, T. M. McQueen, I. A. Troyan, T. Palasyuk, M. I. Eremets, R. J. Cava, S. Naghavi, F. Casper, V. Ksenofontov, G. Wortmann, and C. Felser: 
Nat. Mater. {\bf 8}, 630 (2009). 
\bibitem{Wu1} M. K. Wu, F. C. Hsu, K. W. Yeh, T. W. Huang, J. Y. Luo, M. J. Wang, H. H. Chang, T. K. Chen, S. M. Rao, B. H. Mok, C. L. Chen, Y. L. Huang, C. T. Ke, P. M. Wu, A. M. Chang, C. T. Wu, and T. P. Perng: 
Physica C {\bf 469}, 340 (2009).
\bibitem{Han1} Y. Han, W. Y. Li, L. X. Cao, S. Zhang, B. Xu, and B. R. Zhao: 
J. Phys. Condens. Matter {\bf 21}, 235702 (2009).
\bibitem{Bellingeri1} E. Bellingeri, R. Buzio, A. Gerbi, D. Marr{\`e}, S. Congiu, M. R. Cimberle, M. Tropeano, A. S. Siri, A. Palenzona, and C. Ferdeghinia: 
Supercond. Sci. Technol. {\bf 22}, 105007 (2009). 
\bibitem{Si1} W. Si, Z. W. Liy, Q. Jie, W. G. Yin, J. Zhou, G. Gu, P. D. Johnson, and Q. Li: 
Appl. Phys. Lett. {\bf 95}, 052504 (2009). 
\bibitem{Imai1} Y. Iami, R. Tanaka, T. Akiike, M. Hanawa, I. Tsukada, and A. Maeda: 
Jpn. J. Appl. Phys. {\bf 49}, 023101 (2010).
\bibitem{Bellingeri2} E. Bellingeri, I. Pallecchi, R. Buzio, A. Gerbi, D. Marr{\`e}, M. R. Cimberle, M. Tropeano, M. Putti, A. Palenzona, and C. Ferdeghini: 
Appl. Phys. Lett. {\bf 96}, 102512 (2010). 
\bibitem{Imai2} Y. Imai, T. Akiike, M. Hanawa, I. Tsukada, A. Ichinose, A. Maeda, T. Hikage, T. Kawaguchi, and H. Ikuta: 
Appl. Phys. Express {\bf 3}, 043102 (2010). 
\bibitem{Hanawa1} M. Hanawa, A. Ichinose, I. Tsukada, Y. Imai, T. Akiike, T. Hikage, T. Kawaguchi, H. Ikuta, and A. Maeda: 
to be published in Jpn. J. Appl. Phys.
\bibitem{C1'} This sample is another six-terminal shape film simultaneously 
prepared on the same substrate of film C1 and named C1'. 
\bibitem{Klein1} T. Klein, D. Braithwaite, A. Demuer, W. Knafo, G. Lapertot, C. Marcenat, P. Rodi{\`e}re, I. Sheikin, P. Strobel, A. Sulpice, and P. Toulemonde: 
Phys. Rev. B {\bf 82}, 184506 (2010). 
\bibitem{Tsurkan1} V. Tsurkan, J. Deisenhofer, A. G{\"u}nther, Ch. Kant, H.-A. Krug von Nidda, F. Schrettle, and A. Loidl: 
arXiv:1006.4453v2. 
\bibitem{Iida2} K. Iida, S. Haindl, T. Thersleff, J. H{\"a}nisch, F. Kurth, M. Kidszun, R. H{\"u}hne, I. M{\"o}nch, L. Schultz, B. Holzapfel, and R. Heller: 
Appl. Phys. Lett. {\bf 97}, 172507 (2010).
\bibitem{Lee1} S. Lee, J. Jiang, Y. Zhang, C. B. Bark, J. D. Weiss, C. Tarantini, C. T. Nelson, H. W. Jang, C. M. Folkman, S. H. Baek, A. Polyanskii, D. Abraimov, A. Yamamoto, J. W. Park, X. Q. Pan, E. E. Helstrom, D. C. Larbalestier, and C. B. Eom: 
Nat. Mater. {\bf 9}, 397 (2010). 
\bibitem{Eom1} C. B. Eom, M. K. lee, J. H. Chol, L. J. Belenky, X. Song, L. D. Cooley, M. T. Naus, S. Patnalk, J. Jiang, M. Rikel, A. Polyanskii, A. Gurevich, X. Y. Cal, S. D. Bu, S. E. Babcock, E. E. Hellstrom, D. C. Larbalestier, N. Rogado, K. A. Regan, M. A. Hayward, T. He, J. S. Slusky, K. Inumaru, M. K. Haas, and R. J. Cava: 
Nature {\bf 411} 558 (2001). 
\bibitem{Ikuta1} H. Ikuta, private communication. 

\end{thebibliography}
\end{document}